\documentclass[conference]{IEEEtran}
\IEEEoverridecommandlockouts
\usepackage{cite}
\usepackage{amsmath,amssymb,amsfonts}
\usepackage{algorithmic}
\usepackage{graphicx}
\usepackage{textcomp}
\usepackage{xcolor}
\def\BibTeX{{\rm B\kern-.05em{\sc i\kern-.025em b}\kern-.08em
    T\kern-.1667em\lower.7ex\hbox{E}\kern-.125emX}}

\begin{document}

\title{A 5G NR based System Architecture for Real-Time Control with Batteryless RFID Sensors
\thanks{}
}

\author{\IEEEauthorblockN{Peng Hu }
\IEEEauthorblockA{\textit{Digital Technologies Research Center} \\
\textit{National Research Council of Canada}\\
Waterloo, Canada \\
Peng.Hu@nrc-cnrc.gc.ca}
% \and
% \IEEEauthorblockN{2\textsuperscript{nd} Given Name Surname}
% \IEEEauthorblockA{\textit{dept. name of organization (of Aff.)} \\
% \textit{name of organization (of Aff.)}\\
% City, Country \\
% email address}
% \and
% \IEEEauthorblockN{3\textsuperscript{rd} Given Name Surname}
% \IEEEauthorblockA{\textit{dept. name of organization (of Aff.)} \\
% \textit{name of organization (of Aff.)}\\
% City, Country \\
% email address}
}

\maketitle

\begin{abstract}
The fifth-generation wireless networking (5G) technologies have been developed to meet various time-critical use cases with ultra-reliable, low-latency and massive machine-type communications which are indispensable for tactile Internet applications. Recent advancements in very low-cost and batteryless radio-frequency identification (RFID) sensors have given promises of deploying a massive amount of such sensors for real-time sensing and control applications on a 5G New Radio (NR) network. However, the system design and performance of such applications have not been well studied. This paper proposes a novel system architecture for the representative batteryless RFID touch sensors in generic real-time control applications in a 5G NR mmWave environment. We will discuss the solution using edge computing nodes on the 5G NR base station to the implementation of the proposed system architecture. The real-time performance evaluation with the comparison of the Long-Term Evolution (LTE) networks has shown the effectiveness of the proposed system architecture.

\end{abstract}

\section{Introduction}
The recent advancements of sensing, computing, and wireless networking technologies have given great promises for the tactile Internet. Tactile Internet is ``a network, or a network of networks, for remotely accessing, perceiving, manipulating, or controlling real and virtual objects or processes in perceived real time'' as defined by the IEEE P1918.1 working group, and it has huge potentials to be used in human-to-machine (H2M) and machine-to-machine (M2M) applications \cite{ITU2014}, such as industrial automation, haptic \cite{Antonakoglou2018}, Industry 4.0 \cite{Gundall2018}, and massive Internet of Things (IoT) systems. 

Recently, passive radio-frequency identification (RFID) tags in the ultra-high frequency (UHF) band have been studied for various so-called ``battery-free'' or ``batteryless'' sensing solutions. With the latest batteryless RFID sensors technologies \cite{Bhattacharyya2011, Li2015, Hsieh2018_RFID, Wang:2018} and ultra-reliable and low-latency communication (URLLC) to be fundamentally supported by 5G New Radio (NR) networks in the millimetre wave (mmWave) range, a wide range of H2M and M2M applications meeting the tactile Internet goals can be realized. For example, the batteryless RFID temperature sensor \cite{Bhattacharyya2011}, touch sensors \cite{Li2015, Wang:2018, Hsieh2018_RFID}, and various commercial sensors on the market may generate control or environmental data to be sent to a remote endpoint over a wireless network. Although there are existing works discussing the system architectures of using 5G NR technologies for the URLLC or tactile Internet \cite{Aijaz2019, Antonakoglou2018, Hou2019, Kokkonis2016, Parvez2018}, specific architectures and performance analysis for the tactile Internet applications with the latest batteryless RFID sensors are hardly studied. Furthermore, the challenging question ``how to design a time-critical system for batteryless RFID sensors on a 5G NR network'' is yet to be answered.

In this paper, we will address this question through the discussion on the system architecture for the real-time control applications enabled by the new batteryless RFID touch sensors \cite{Li2015, Wang:2018}, where the RFID sensors can respond to various touch events and harvest energy from radio frequency (RF) signals of an ambient RFID reader. With the proposed 5G based architecture, the deployment of such sensing applications can benefit from the enhanced flexibility and scalability that extends current local setups of typical RFID sensing systems. For example, touch sensors can be used not only in building automation systems but also in typical tactile Internet applications without wired connections. 

In summary, the following contributions are made:
\begin{itemize}
    \item We have proposed a new system architecture for batteryless RFID touch sensors which significantly extends their applications in various use cases. The key proposed elements in the proposed architecture are identified.
    \item We have proposed a new solution based on the 5G NR network and edge computing (EC) to realizing the implementation of the proposed system architecture. 
    \item We have evaluated the real-time performance with the comparison of the classical Long-Term Evolution (LTE) networks and showed the superior effectiveness of the proposed system architecture.
    \item We have identified the considerations when deploying a massive number of RFID sensors in a real-world scenario.
\end{itemize}

The rest of the paper is structured as follows: Section II briefly presents the related work; Section III discusses the proposed system architecture, its key architectural elements and protocol stack; Section IV discusses the performance evaluation of the system setups following the proposed system architecture deployed on LTE and 5G NR networks; and the conclusive remarks are made in Section V.

\section{Related Work}
The tactile Internet has recently been coined out as an attractive paradigm for real-time sensing and control in a broad range of application domains. Recent studies on batteryless RFID sensors with different sensing capabilities have enabled many attractive solutions to tactile Internet. Basically, a batteryless RFID sensor is a passive ultra-high-frequency (UHF) RFID tag with an antenna and a chip. When designed as a sensor, its antenna impedance can change with a correlation to various physical properties, such as touch, pressure, moisture, temperature, etc. For example, a low-cost and single-use RFID sensor in \cite{Bhattacharyya2011} is designed to check the temperature threshold where the violations of temperature are converted to a change of antenna properties of the UHF RFID tag. The IDSense system \cite{Li2015} with battery-free RFID tags has been proposed to sense different motion and touch events with commercial UHF RFID tags. The RFIBricks system \cite{Hsieh2018_RFID} provides a new interactive human-machine interface using custom UHF RFID tags on blocks for detecting various user inputs. The RFID sensor designs for sensing light, temperature, touch, and gesture events are discussed in \cite{Wang:2018}. Theses batteryless RFID sensing solutions are promising; however, the limitations of the solutions including the short communication range between an RFID reader and an RFID tag/sensor need to be overcome for a broader range of applications in the tactile Internet.

The enabling framework and networking technologies for tactile Internet have also been advancing in recent years. The IEEE P1918.1 working group is working on the framework of the tactile Internet in the multi-disciplinary context because of its broad use cases. One typical use case is haptic applications, where Kokkonis \textit{et al.} \cite{Kokkonis2016} discuss the performance metrics in terms of the round-trip time (RTT) and packet loss ratio (PLR) in a local IEEE 802.11g network and show that the performance for transmitting small packets can be decreased with the use of an access point. Ajaz and Sooriyabandara \cite{Aijaz2019} discuss the network architecture with the use of 5G NR technologies for the tactile Internet.

5G is expected as a strong enabling technology for tactile Internet applications. Antonakoglou \textit{et al.} \cite{Antonakoglou2018} discuss the communication options for 5G-based haptic communication transmission using networked teleoperation systems. Moutaser \textit{et al.} \cite{Mountaser2018} present the design of fronthaul unit in a Cloud-RAN environment, where a fronthaul unit is located between the central unit and radio unit supporting various radio interfaces. Mahlouji and Mahmoodi \cite{Mahlouji2018} discuss the use of four uplink scheduling schemes in haptic traffic. To achieve an efficient allocation of radio resources, Hou \textit{et al.} \cite{Hou2019} propose a prediction scheme based on future system states which can be sent to the receiver for minimizing the communication latency for URLLC. A coordinated multi-cell resource allocation targeting at minimizing the inter-cell interference was proposed in \cite{Hytonen2017} for the indoor 5G URLLC solutions. Parvez \textit{et al.} \cite{Parvez2018} survey the latency requirement for typical application domains, where the factory automation, gaming, and teleoperations have the latency requirements as low as around 1 ms although the data amount may vary. One of the important requirements when using touch sensors for tactile Internet applications is timing guarantees in terms of the latency, which can be defined as the difference between the time when a touch event on touch sensors occurs and the time an object is controlled by such an event. The aforementioned works, however, have not specifically targeted at the batteryless RFID sensors, and the study of a working system architecture for batteryless RFID sensors in a 5G NR environment is lacking.% For typical H2M applications, we need to keep the latency within 50 ms for good user experience. For time-critical industrial M2M applications, we need to keep the latency within a few milliseconds to a few hundred milliseconds \cite{Palattella2016}.

\section{System Architecture}
\subsection{A Typical Setup of Batteryless RFID Touch Sensor Systems}
\begin{figure}[ht]
\centering
\includegraphics[width=2.8in]{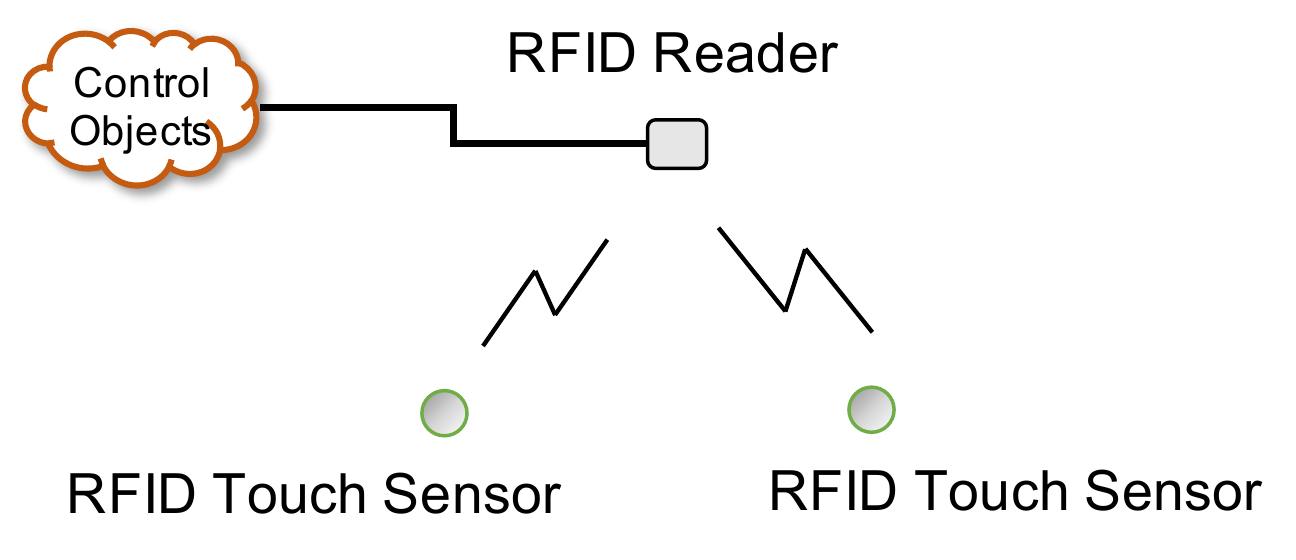}
\caption{A typical system setup of a control application using an RFID reader and batteryless RFID touch sensors.}
\label{classical_setup}
\end{figure}

A typical system setup for a control application using batteryless RIFD touch sensors is shown in Fig. \ref{classical_setup}, which consists of an RFID reader (compliant with the ISO/IEC 18000 standard), a set of RFID touch sensors, and a wired connection to control objects. In Fig. \ref{classical_setup}, the RFID reader is in proximity to the RFID touch sensors in order to emit radio-frequency (RF) signals in the UHF band to power the RFID touch sensors and to receive the RF signals responded from the sensors. After the received signals are processed at the RFID reader, commands can be issued to control the objects connected. 

We can see that this system setup has several limitations. First, the flexibility of the system is poor as RFID reader, sensors and control objects usually need to be located in proximity to each other and it can only control objects through a wired connection. Second, the compatibility of the system is limited. For example, for industrial automation, if the control objects are based on the Controller Area Network (CAN), it requires the matching CAN module on the RFID reader. Last but not least, the number of touch sensors supported is dependant on the capability of the RFID reader used which limits its scalability. 

\subsection{A Novel 5G NR based System Architecture}

\begin{figure}[ht]
\centering
\includegraphics[width=3.5in]{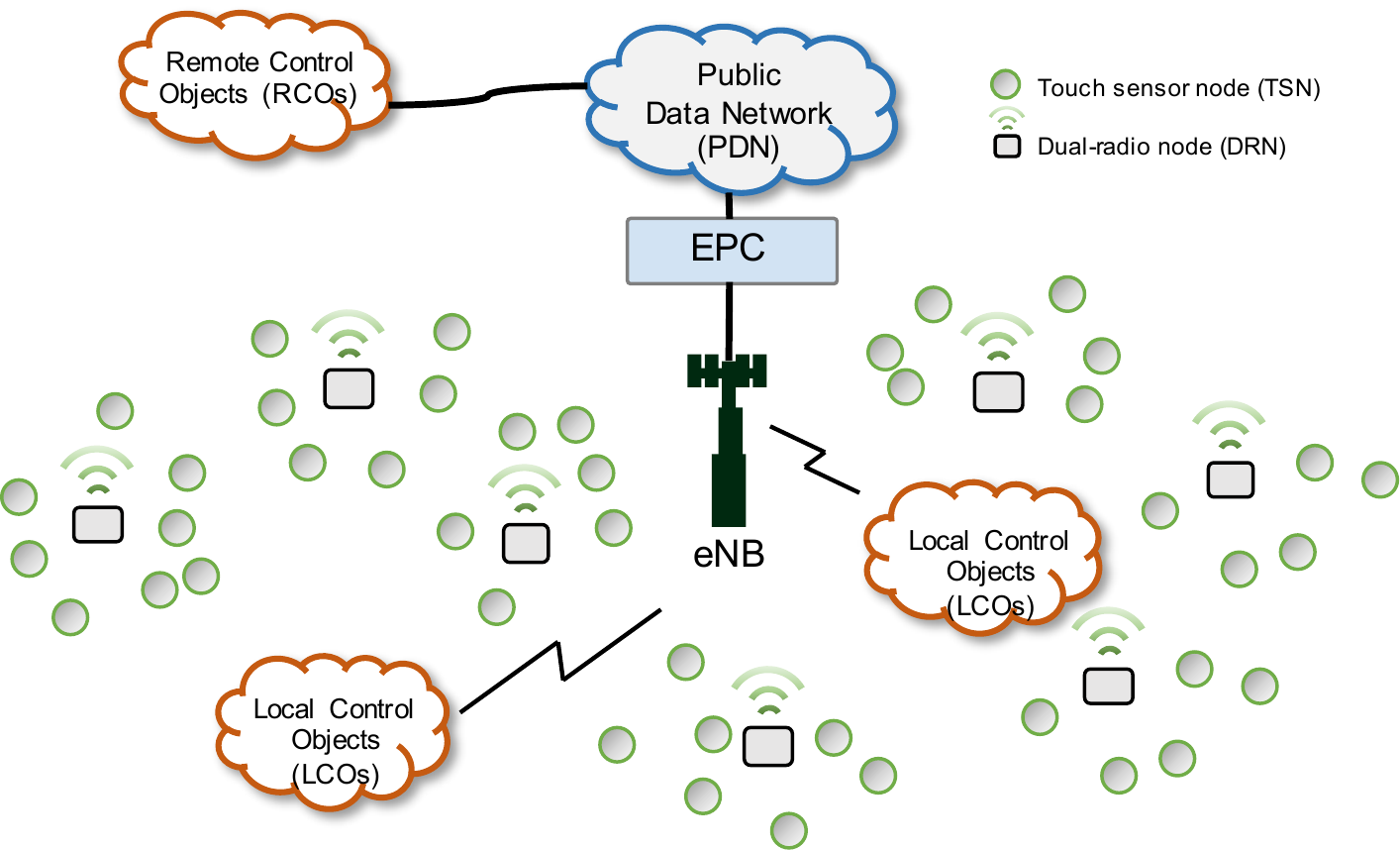}
\caption{The proposed system architecture consisting of the cellular network and key elements such as TSNs, RCOs, LCOs, and DRNs, where each DRN is equipped with an RFID reader module and a cellular UE module.}
\label{system_arch}
\end{figure}

We will consider the system architecture with a cellular base station (BS) or evolved NodeB (eNB) as well as a number of RFID touch sensors shown in Fig. \ref{system_arch}. This general setup can work in many use cases, such as building automation and industrial automation. The key elements are defined as follows:

\begin{itemize}
    \item A touch sensor node (TSN) is a batteryless RFID touch sensor powered by an adjacent RFID reader. 
    \item A dual-radio node (DRN) is a device equipped with an RFID reader and the user equipment (UE) module which can interact with TSNs and co-located eNBs. There may be multiple geographically distributed DRNs, where each DRN is associated with local TSNs in a system as shown in Fig. \ref{system_arch}.
    \item A local control object (LCO) is an object to be locally controlled via one or more co-located eNBs as part of the local radio access network (RAN). There may be multiple LCOs in a system.
    \item A remote control object (RCO) is an object to be remotely controlled via communication networks which are beyond the coverage of co-located eNBs. Multiple RCOs or a combination of RCOs and LCOs may be present in a system.
\end{itemize}

In addition, an LCO contains a UE module and can be controlled with the control data originated from a touch event on a TSN via the elements such as DRN and eNB. Similarly, an RCO contains a UE module and other network interface modules, which can be controlled via public data networks (PDNs) and the local wireless network. The proposed system architecture can relax the limitations in the typical system setup at least in the following factors:
\begin{itemize}
    \item Flexibility is enhanced as the control objects are free from wired connections to the local RFID system.
    \item Scalability is enhanced as multiple LCOs deployed in a space such as a business building can enjoy the existing coverage by a RAN.
    \item Compatibility is enhanced as the LCOs or RCOs can use the standard wireless radio interface for taking the commands with backward and forward compatibility. Also, the DRN allows for a gateway implementation for translating different protocols.
\end{itemize}

\subsection{Protocol Stack Architecture for Key Elements}

\begin{figure}[ht]
\centering
\includegraphics[width=3.5in]{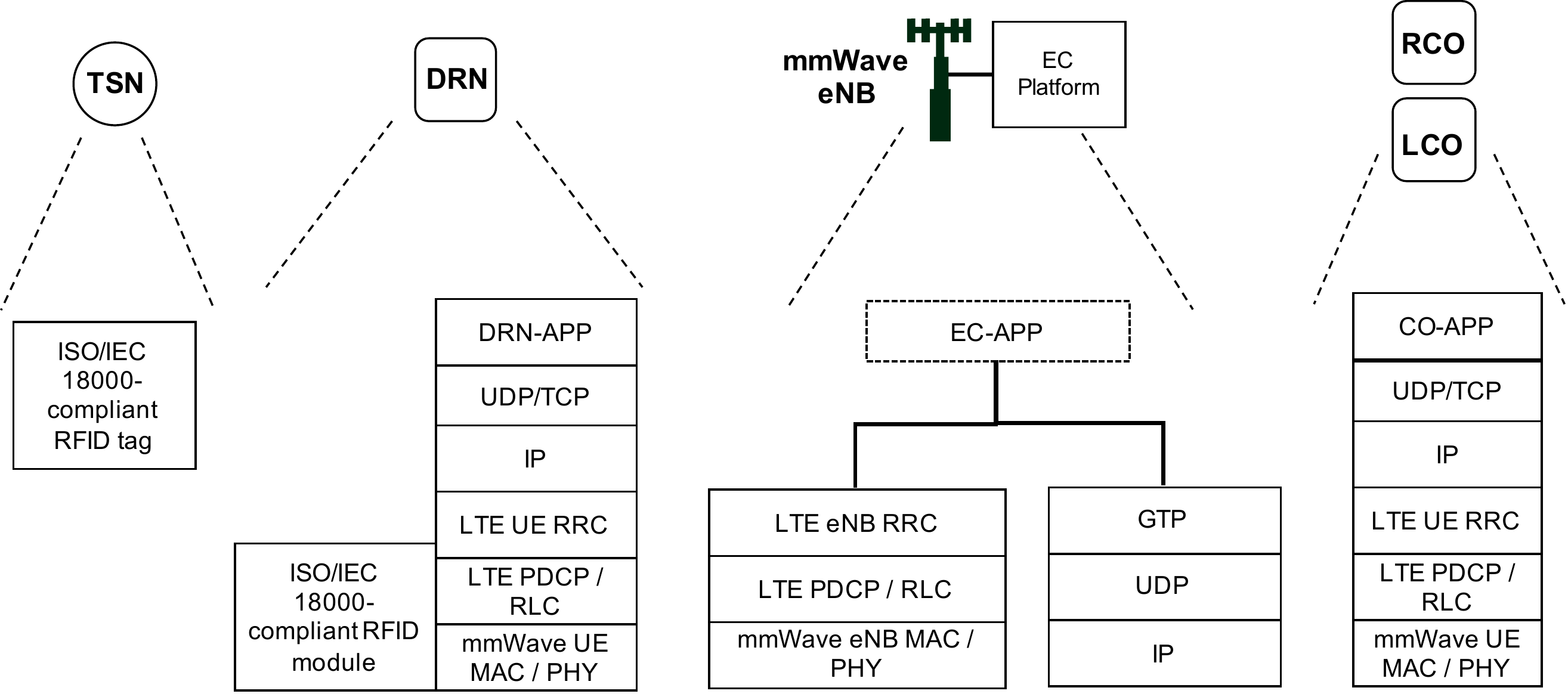}
\caption{Protocol stack architecture for key architectural elements in the proposed system architecture.}
\label{stack}
\end{figure}

The protocol stack architecture for the key elements is shown in Fig. \ref{stack}, where both TSN and DRN contain a standard radio interface compliant with ISO/IEC 18000 standard, and the DRN also contains a full stack of a mmWave air interface as the standard user equipment (UE), including the entities such as physical layer (PHY), medium access control (MAC) sub-layer, packet data convergence protocol (PDCP), radio link control (RLC), radio resource control (RRC), Internet protocol (IP), user datagram protocol (UDP), transmission control protocol (TCP), and an application (APP) endpoint. The mmWave eNB node shown in Fig. \ref{stack}, which may be connected to an EC platform running an EC application (EC-APP) endpoint, contains the layer-1 and layer-2 mmWave air interface protocols as well as the essential IP-based protocols such as IP, UDP, and General Packet Radio Service (GPRS) tunneling protocol (GTP). These protocols are essential for interconnections with the evolved packet core (EPC) entities such as packet gateway (P-GW) and IP-based hosts on a public data network (PDN). On the LCO or RCO node, there is a standard UE stack and IP stack with a control object application (CO-APP) endpoint interacting with a DRN application (DRN-APP) endpoint.

The proposed stack architecture can work with the 5G NR system of various deployment options. 

\subsection{Discussion on Design Considerations}

First, we should note that there are several general assumptions implied in the proposed system architecture and its elements, which are listed as follows:
%First, the immobility of TSNs and DRNs are mostly immobile and the channel conditions are assumed quasi-static. Second, the data from $N_{t}$ TSNs can reasonably be considered to be events generated. This means in some cases, the events that occur before time $t$ are random variables (RVs) following a stochastic process, such as a Poisson process. In this case, we can simulate the TSN data based on exponential RV. Third, transferring TSN data can have different policies on the key local elements such as DRNs, eNBs, RCOs and LCOs. Fourth, two major performance metrics are end-to-end latency and reliability of the transmission. Fifth, from the networking perspective badsed on Fig. \ref{system_arch}, the RCOs and LCOs can be jointly considered in the modelling, which means RCOs can be considered as remote LCOs when additional entities such as EPC and PDN entities are added between a DRN and an LCO. At last, the RFID reader radios on DRNs are assumed to have sufficient separation between each other to avoid RF interference. 
\begin{itemize}
    \item TSNs and DRNs are mostly stationary and the channel conditions are assumed quasi-static.
    \item RFID reader radios on DRNs are assumed to have good separation from each other to avoid interference.
    \item TSN data may be transferred with different policies on the key elements.
    \item Data from TSNs can be considered to be events generated.
    \item RCOs and LCOs can be jointly considered in system modelling. For example, RCOs can be viewed as remote LCOs when additional elements such as EPC and PDN entities are added between a DRN and an LCO.
\end{itemize}

An important requirement when using TSNs for tactile Internet applications is the timing guarantee in terms of the latency $T_G \leq \epsilon, \epsilon \in \mathbf{R}^+$, where $\epsilon$ is the threshold and $T_G$ is the difference between the time when the touch event on touch sensors occurs and the time an object is controlled by the event. For consumer IoT (CIoT) applications, we may need to keep the latency within 50 ms for a good user experience. For time-critical industrial IoT (IIoT) or Industry 4.0 applications such as factory automation, we need to keep the latency within a few milliseconds to a few hundred milliseconds \cite{Palattella2016}. 

In the context of EC, an eNB in the LTE and 5G network is considered to have a set of limited compute resources ($R_{C}$) and a set of communication resources in terms of resource blocks ($R_{RB}$). With these resources, we need to ensure the latency can be achieved. The latency here is considered as end-to-end data transmission latency $T_D$, which, as the dominant part of $T_G$, is the time difference between when a data packet is sent from the TSN and when the data is received at an LCO or RCO. %The problem now is to ensure an optimal allocation policy $\pi$ of $R_{RB}$ which can meet the latency of data transmission within $\epsilon$, where we assume the data transmission latency $T_D$ is the dominant part of $T_G$, and $T_D$ is defined as the time difference between when touch sensor data is emitted from the touch sensor and when the data is received at the controlled object. 

Although the parameters can be well considered with the proposed architecture, the internal modules for each key element such as internal computation and resource scheduling modules at various layers of a node are beyond the scope of this paper.

\subsection{Using the Proposed Architecture in Generic Applications}

\begin{figure}[ht]
\centering
\includegraphics[width=3.52in]{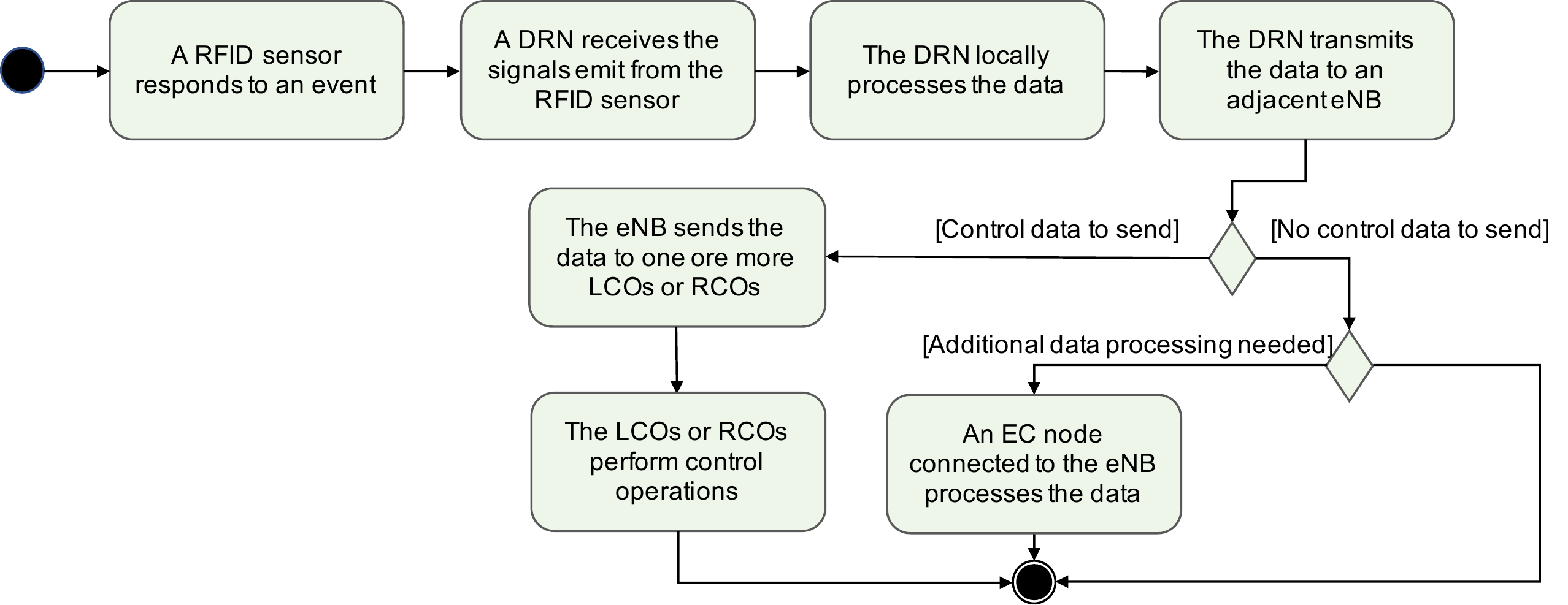}
\caption{An activity diagram for generic real-time sensing and control applications based on the proposed architecture.}
\label{activity_diagram}
\end{figure}

Now let us see how the proposed system architecture can help with a generic real-time sensing and control application. Fig. \ref{activity_diagram} shows an activity diagram for a generic real-time control application. The first activity in Fig. \ref{activity_diagram} involves the event sensing specific to a type of batteryless RFID sensors, which may take a variable amount of time for a sensing event due to the sensing mechanism of the RFID sensors. The remaining activities can generally be addressed with the key architectural elements, where DRNs play a key role in ensuring the real-time data transmission to LCOs and RCOs. An EC node can be used in additional processing of local sensing data sent from DRNs. The activities shown in Fig. \ref{activity_diagram} indicate the places for optimization if we want to use the proposed architecture for a specific application. For example, in an industrial monitoring application using batteryless RFID temperature sensors, for a given $\epsilon$, we may optimize the sensing data transmissions with DRNs and eNBs using the limited $R_C$ and $R_{RB}$. The sensing data packets from an RFID sensor and data flows between multiple DRNs and eNBs and between eNBs and LCOs/RCOs can be optimized to satisfy an end-to-end timing guarantee between the sensor and the LCO or RCO.

% Use the typical numerology with the load in terms of N_t and N_d
% Show we can overcome the cons by using adaptive strategy at the MAC controller level

%We can formulate the optimization problem as follows:
%The latency of the LTE link can be calculated through the 

\subsection{Discussion on Real-World Use Cases}
The proposed architecture can be used in general real-time sensing and control applications based on generic batteryless RFID sensors including touch sensors. Let us briefly summarize them in a few examples here.

\subsubsection{Building Automation} The building automation is a typical use case where the proposed architecture can be used. For example, in a building management system, some control panels as LCOs and DRNs associated with batteryless RFID sensors can be programmed for various control functions, such as lighting, heating, ventilation, and security, where the physical devices as LCOs are distributed across a building.

\subsubsection{Factory Automation} A factory automation system demands real-time control for various automation tasks, which can be provided with the proposed architecture. Although the radio environment in an industrial environment is generally complex, it is possible to employ a well-designed system based on the proposed architecture to perform various sensing tasks of the automation lines with batteryless RFID sensors.

\subsubsection{Environmental Monitoring} Batteryless RFID sensors having capabilities of sensing environmental data, such as temperature, air quality, and radiation, can be deployed in a distributed fashion in different areas indoors or outdoors to enable short-term or long-term environmental monitoring applications, where a long-term environmental monitoring application can enjoy the maintenance-free and passive characteristics of the RFID sensing system. 

\subsubsection{Haptic and Medical} As a typical use case aligned with the tactile Internet, we can use the batteryless RFID touch sensors to enhance the current haptic and medical applications with the sensing capability for touch events. 

\subsubsection{Interactive Gaming and Virtual Reality} An interactive gaming system in a virtual reality environment using an RFID touch sensor, such as RFIBricks \cite{Hsieh2018_RFID}, can be achieved by the proposed architecture on a 5G NR network to facilitate the real-time transmission of the gaming and touch sensing data.

The aforementioned use cases are non-exhaustive. With the extension of the DRN for a specific type of batteryless RFID sensors, one can easily adapt the proposed architecture for different applications.

%\subsubsection{Structural Health Monitoring} RFID sensors with pressure or strain sensing capability can be employed to 

\section{Performance Evaluation}

Here we evaluate the essential end-to-end real-time performance based on the proposed system architecture in the previous section, where simulations of a network with one eNB and a various number of DRNs and LCOs are performed. The ns-3 LTE and mmWave modules \cite{Mezzavilla2018} are used in the simulations, where the mmWave module utilizes the standardized beamforming and modulation coding scheme (MCS) level to determine the transport block size and subframe slots based on the 3GPP channel model with central frequency at 28 GHz. More specifically, the standard 1-ms 14-symbol subframe is used on the LTE network and a flexible transmission time interval (TTI) scheme with the 100-$\mu$s 24-symbol subframe period \cite{Mezzavilla2018} is used on the mmWave network. We evaluate the real-time performance of the system setups based on the proposed architecture using an LTE backhaul and a mmWave backhaul, respectively. The TSNs are assumed to be pre-associated to a DRN and the data from TSNs are transmitted to LCO through an eNB, which involves the uplink (UL) and downlink (DL) connections. The transmission of sensing data from a DSN to an eNB is via a UL connection while that data gets processed at the EC platform on an eNB and transmit from the eNB to an LCO via a DL connection. IPv6 is adopted on all nodes and the DRN and LCO nodes are randomly deployed on a 100 m $\times$ 100 m plane.

A client-server UDP application is used on DRN-APP, EC-APP, and CO-APP entities, where the DRN-APP is the client and LCO-APP is the server, where the EC-APP entity simply forwards data. The precise end-to-end latency is tracked through the use of byte tags in ns-3. In the baseline scenario, the application data payload with the size of 1024 B is used. Five runs are performed in each iteration with a random deployment of DRN-LCO pairs for each run.

\begin{figure}[ht]
\centering
\includegraphics[width=3.5in]{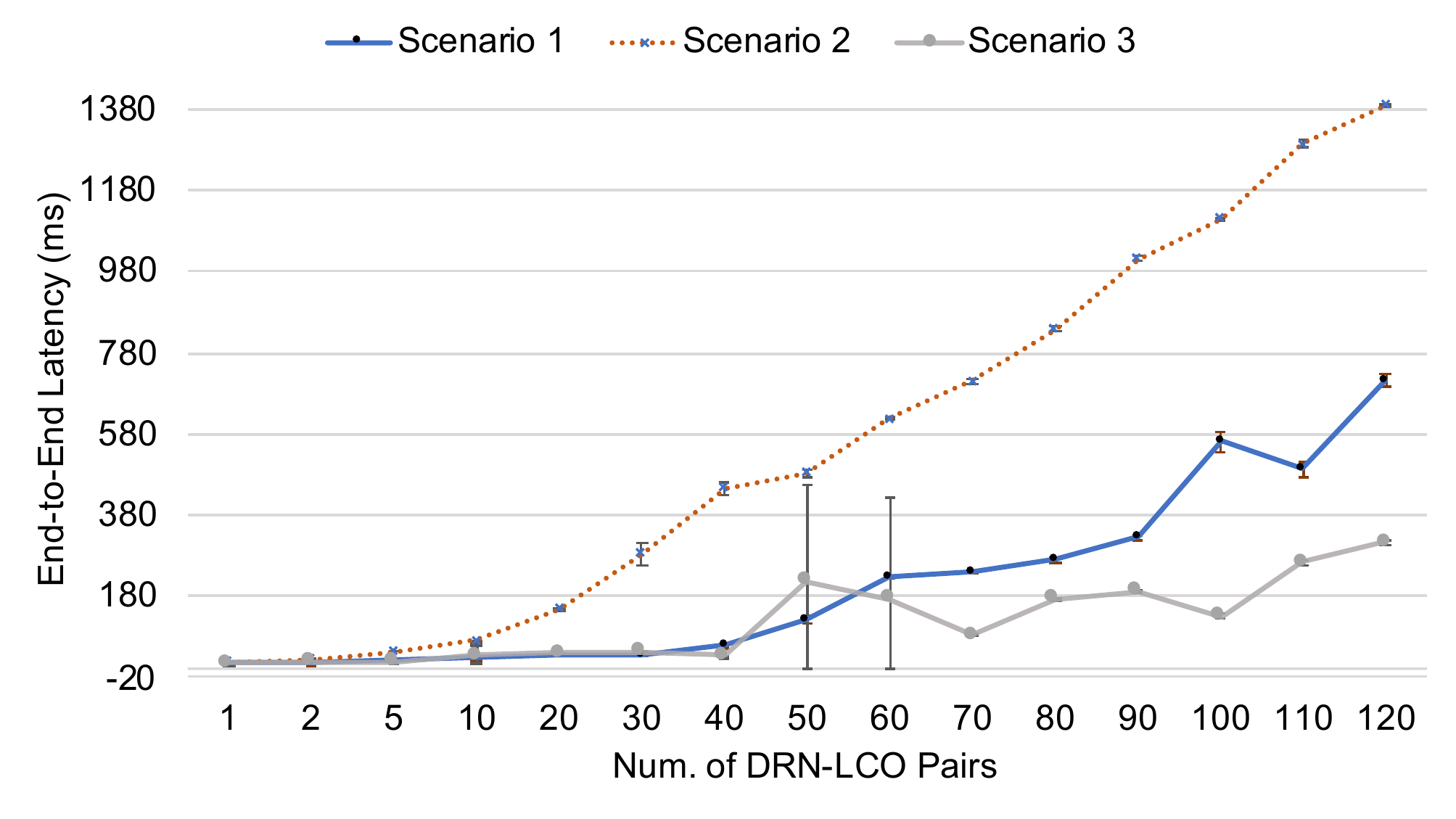}
\caption{The performance of average end-to-end latency between DRN and LCO versus number of DRN-LCO pairs in an LTE network shown in Fig. \ref{system_arch}. }
\label{lte_performance}
\end{figure}

First, based on the network setup shown in Fig. \ref{system_arch}, we first evaluate the performance of the system deployment in the LTE network in three scenarios. In the baseline Scenario 1, each DRN is forwarding the 1024 B data payload received from TSNs to the eNB node and then eNB node immediately transmits the packet received to the LCO. The results are shown in Fig. \ref{lte_performance}, where we can see that with the increasing number of DRNs with TSNs, the overall reliability and latency get degraded over time. The average end-to-end latency $T_D$ (referred to as latency afterwards) ranges from 14.4 ms (when the number of DRN-LCO pairs is 1) to 713.55 ms (when the number of DRN-LCO pairs is 120). Such latency performance can rule out its applications in various real-time systems. This is due to several factors, such as sub-optimal radio resource allocations on LTE eNB, interference, and computation load on the key local entities. From Fig. \ref{lte_performance}, when the number of DRN-LCO pairs is greater than 90, the latency performance gets noticeably worse as more nodes cause additional conflicts in the radio resource scheduling and allocation processes. In Scenario 2 and Scenario 3, each DRN node forwards 5120 B payload and 512 B payload, respectively. We can see the latency performance in Scenario 2 is mostly scaled compared to that of Scenario 1 when the number of DRN-LOO pairs $\geq 2$. In Scenario 3, the latency performance is less than that of Scenario 1 when the number of DRN-LCO pairs $\geq 70$. In all scenarios, we can see the minimum values of the latency are similar.

%For the user control applications, in addition to the timing requirement, the reliability is also important. This means that the any data generated by user operations on TSN should be delivered to the LCOs and RCOs with very high probability. We will see how such data will be allocated with different performance in a mmWave network with an eNB following Fig. \ref{system_arch}.

\begin{figure}[ht]
\centering
\includegraphics[width=3.5in]{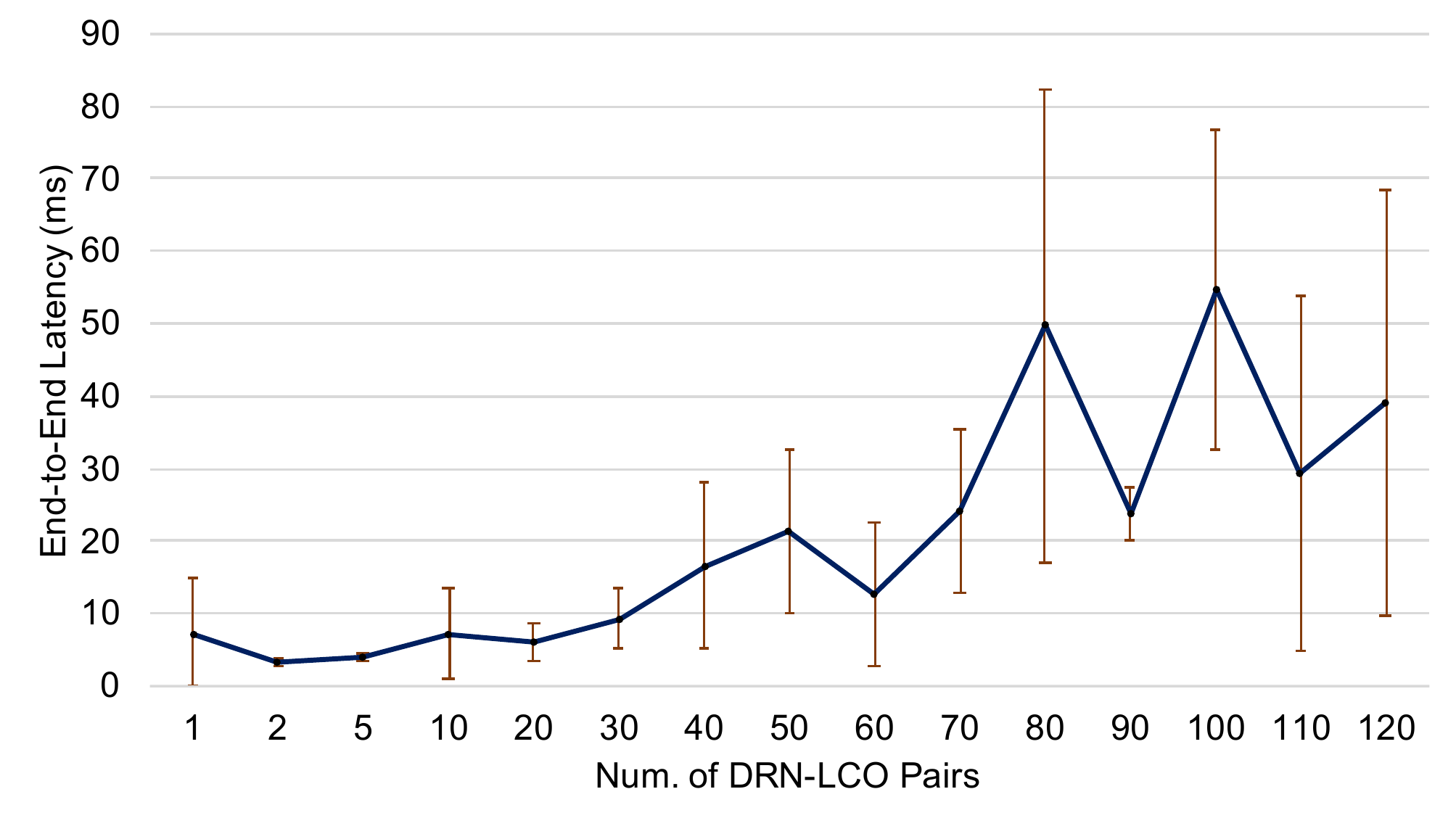}
\caption{The performance of average end-to-end latency between DRN and LCO versus number of DRN-LCO pairs on a 5G NR network}
\label{FigMmWave}
\end{figure}

%Show the mmWave performance
Now let us see the performance of the system in the baseline Scenario 1 on a 5G NR network with one mmWave eNB and various mmWave-based DRNs and LCOs. In Fig. \ref{FigMmWave}, we can see that the end-to-end latency is much lower than that in Fig. \ref{lte_performance}. The overall latency can be well kept within around 60 ms, and when the number of DRN-LCO pairs is less than 30, the latency on average can be maintained within 10 ms. The latency increases a bit when the nodes increase due to the limited resources and the increasing probability of conflicts.

\begin{figure}[ht]
\centering
\includegraphics[width=3.5in]{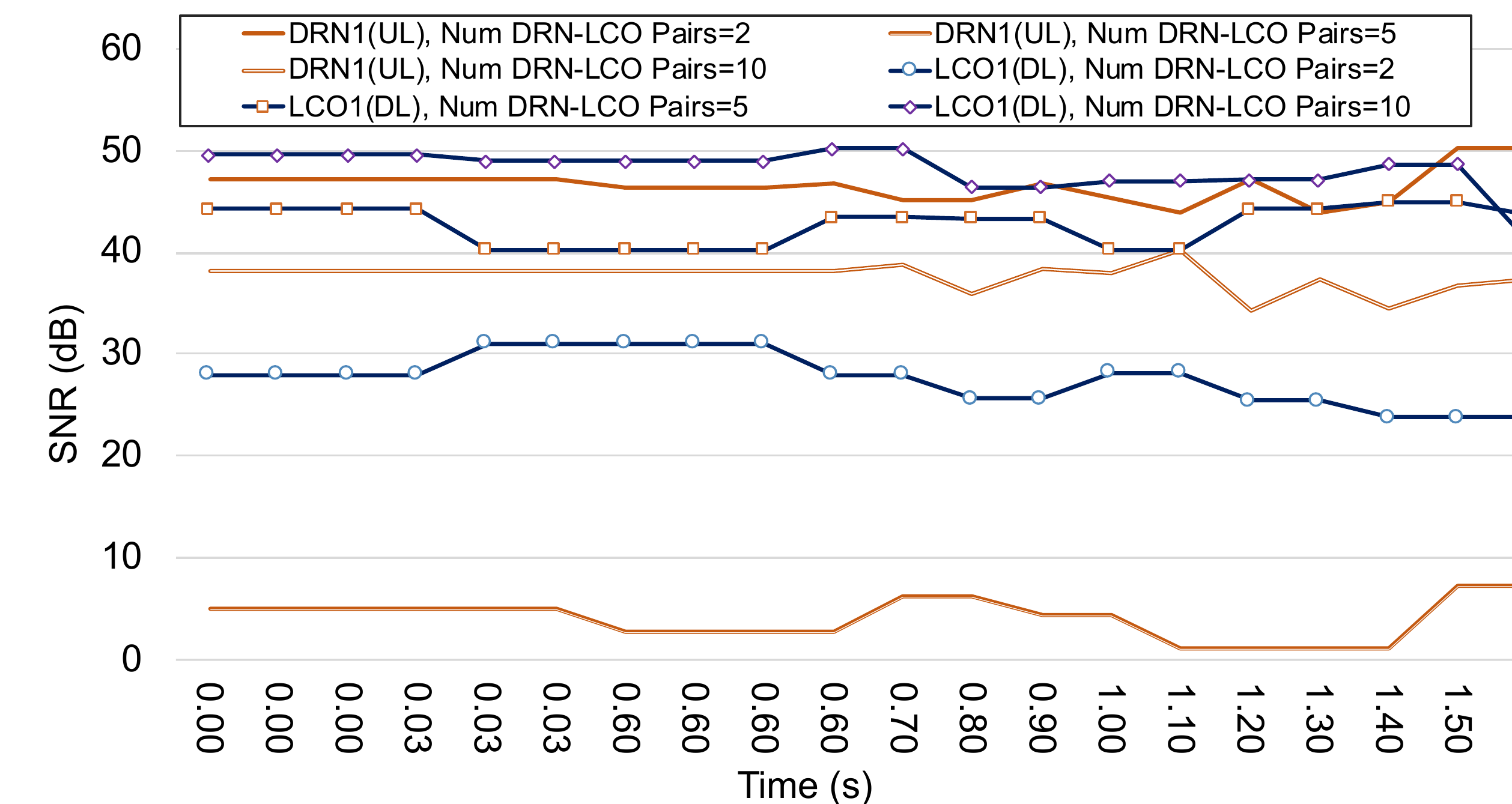}
\caption{SNR condition on DRNs and LCOs over time in three iterations with different network sizes.}
\label{FigSimuSNR}
\end{figure}

The latency performance has shown some fluctuations in \ref{FigMmWave} where the network size in terms of the number of DRN-LCO pairs changes. This is mainly due to the random position generation of the DRN and LCO nodes, which partially changes the channel conditions as shown in Fig. \ref{FigSimuSNR}. In Fig. \ref{FigSimuSNR}, the signal-to-noise ratio (SNR) conditions over time for the DRN/LCO nodes in different network sizes are shown. For each combination of DRN-LCO pairs, the time-series SNR values of the first DRN node (i.e., DRN1) and the first LCO node (i.e., LCO1) during data transmissions are shown in two curves, respectively. The UL of DRN1 and DL of LCO1 are plotted because they are the actual links of transmitting sensing data from the DRN1 end to the LCO1 end. The results in Fig. \ref{FigSimuSNR} reveal the fact that a network planning for a specific network size in a real-world scenario needs to be made on a 5G mmWave network.

In addition, it is observed that the packet size can affect real-time performance. A smaller packet size (e.g., less than 1024 B) sent from a DRN can reduce the latency further, but considering the cases that a DRN may aggregate data from various TSNs, it may need to transfer a large packet. However, transferring a large packet, which basically needs to take a fair amount of communication resources for UL and DL transmissions, requires additional considerations in order to keep the latency within a threshold $\epsilon$.

\section{Conclusions}
The proposed networked system architecture for real-time applications arising from the batteryless RFID touch sensors has many advantages in terms of cost, installation, flexibility, scalability, and compatibility. With the proposed key architectural elements using 5G NR mmWave technologies, real-time control over various objects can be achieved. Through the experimental results in a generic end-to-end sensing and control application, the effectiveness of the proposed system architecture is evaluated, although the latency performance of the mmWave system is susceptible to the radio environment. The proposed architecture can be used as a reference design for enabling many time-critical IoT applications in the tactile Internet and Massive IoT. However, when using the 5G NR as a backhaul network, we need to make sure the deployment of mmWave eNBs is carefully planned, which otherwise would result in poor SNR conditions. Also, the heterogeneity of the various types of RFID sensors on a 5G NR network based on the proposed architecture needs to be studied in the future work.

% \begin{table}[hbt]
% %\renewcommand{\arraystretch}{1.3}
% \caption{Simulation Parameters}
% \label{Tb2:SimuParameter}
% \centering
% \begin{tabular}{c | c}
% \hline
% \textbf{Parameter} & \textbf{Value}\\ 
% \hline
% Bandwidth & (20 MHz, 1 GHz)\\
% Deployment area & $100 \times 100 \text{~m}^2$ \\
% Symbols per subframe  & 24 \\
% Symbol period & 4.16 $\mu${s} \\
% Subframe period & 100 $\mu${s} \\
% RLC Tx Buffer Size & 20 MB \\
% S1 latency & 1 ms \\
% \end{tabular}
% \end{table}

\bibliographystyle{IEEEtran}
% argument is your BibTeX string definitions and bibliography database(s)
\bibliography{IEEEabrv,./main}

\end{document}